\documentclass[12pt]{article}
\usepackage{amsmath,amsthm,amsfonts,graphicx,epsfig}
\usepackage[usenames]{color}

\newcommand{\dbar}{\kern-.1em{\raise.8ex\hbox{ -}}\kern-.6em{d}}

\def \be{\begin{equation}}
\def \ee{\end{equation}}
\begin{document}

\title{The Square Cat} 
\author{E. Putterman and O. Raz
\\
Department of Physics\\ Technion, 32000 Haifa, Israel}

\date{\today}%
\maketitle 
\begin{abstract}

We present a simple, two dimensional example of a "cat" -- a body
with zero angular momentum that can rotate itself with no external
forces. This model is used to explain why this problem is known to
be a gauge theory and to illustrate the importance of
non-commutative operators. We will also show a comparison between
the free-space "cat" in Newtonian mechanics and the same problem in
Aristotelian mechanics at low Reynolds number; this simple example
shows the analogy between (angular) momentum in Newtonian mechanics
and (torque) force in Aristotelian mechanics. We will end by
pointing out a topological invariant common to our model in free space
and at low Reynolds number.
\end{abstract}
\section{Introduction}
It is well known that a cat, falling from a tree with its feet
facing upwards, can rotate itself in midair in order to land on its
feet, even when its net angular momentum is zero. Reorientation of
deformable bodies with zero angular momentum (which we call "cats")
is also important for satellites, astronauts, dancers, divers
\cite{GeometricPhases,ZeroAngTurns,Divers}, and nanomechanics
\cite{OptimalRotation}. At first glance, rotation with zero angular
momentum seems to be an impossible task. This is because we usually
study the angular momenta of rigid bodies, which indeed cannot
rotate with zero angular momentum. Deformable bodies, however, can
reorient themselves with no net angular momentum, as we will show.
Besides being counter-intuitive and interesting,
the rotations of "cats" were shown to be a gauge theory problem \cite{GeometricPhases,Lopez-gauge,Montgomery}, and to be deeply connected to swimming in curved space \cite{TheBaron,Wisdom}.

The canonical model of a falling cat (Kane and Scher
\cite{Kane&Scher}) is two identical axi-symmetric rigid bodies,
connected by a 'no twist' joint. Here we propose a different
model. Our model, which we describe in section
\ref{section:Model}, does not look like a real cat at all; however,
it is manifestly two dimensional, and hence simple to analyze, while
demonstrating the same basic phenomenon of rotating with zero
angular momentum around an axis.

Deformable bodies, besides being able to rotate
with zero angular momentum at free space, are also able to swim
when placed in fluid.  At low Reynolds number, where inertia is
negligible, swimming of deformable bodies shares much with the
reorientation of deformable bodies in free space: the two problems
are known to be gauge problems \cite{GeometricPhases} (we will
discuss this in the free space case in sections
\ref{section:gauge} and \ref{sec:diffGauge}), and in both problems
the ability to change orientation depends on
non-commutative operations (we will discuss this in the case of rotations in
section \ref{sec:non-commute}). We will show in section
\ref{sec:CatInHoney} that our model, at a particular limit, will rotate identically
in free space and at low Reynolds number. In the general case, the model will
generally behave differently in free space and at low Reynolds
number; however, the maximal rotation due to the largest possible stroke must be
the same - as a consequence of a topological invariant which we will
discuss in section \ref{sec:TopoInv}.


\section{The Model}\label{section:Model}
\begin{figure}
  \includegraphics[width=10 cm]{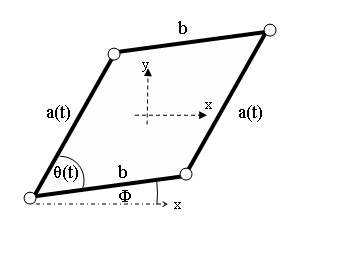}\\
  \caption{\textit{Example 1--} The cat model, $\phi$ chosen as the angle between $x$ and the $b$ rod.}\label{fig:Exmp1}
\end{figure}
The model we propose is composed of 4 spheres, each with mass $m$,
connected by 4 massless rods in the shape of a a parallelogram. The
cat can control the base angle, which we will denote by $\theta$,
and the length of one of the pairs of parallel rods, which we will
call $a$.  The other pair of rods has fixed length which we denote
by $b$ (see Fig. \ref{fig:Exmp1}). Since the model is made of 4
masses, each with 2 degrees of freedom, it has 8 degrees of freedom.
However, there are 5 constraints in this problem: 4 distances
between the masses due to the rods, and one angle. Thus, there are
only three physical degrees of freedom, which can be expressed as 2
degrees for the location of the center of mass, and one degree for
the total rotation. Since it is clear that without external forces
the velocity of the center of mass  will not change, we will work in
the reference frame in which the center of mass is at rest, and we
will not consider its two degrees of freedom. This leaves us with
only one physical degree of freedom - the total rotation of the
body, which we denote by $\phi$. There are many ways to choose the
angle $\phi$, as we will discuss in section \ref{section:gauge};
here we will choose $\phi$ as the angle between the $x$-axis of an
arbitrarily oriented reference frame whose origin is at the center
of mass and one of the $b$ rods (see Fig. \ref{fig:Exmp1}).

We would now like to show that the "cat" can rotate itself by
changing $a$ and $\theta$. The equation of motion of this system is
conservation of angular momentum (which in a 2-dimensional system is
a scalar), $\frac{dL}{dt}=0$. The initial angular momentum is 0, so
this equation is equivalent to ${L} = 0$. We can write $L$ in terms
of the change in $a,\theta,\phi$ - that is, in terms of $\dot{a},
\dot{\theta}, \dot{\phi}$. A direct calculation shows that the
equation of motion is:
\begin{equation}\label{eq:FreeSpaceCat-EqOfMotion}
{L}=4\dot{\phi}(a^2+b^2)+4\dot{\theta}a^2=0
\end{equation}
From this equation we can derive an equation for the change in
$\phi$: since $\dot{\phi}=-\frac{\dot{\theta}a^2}{a^2+b^2}$, we can
write $\Delta\phi=\int{-\frac{\dot{\theta}a^2}{a^2+b^2}dt}$. This
equation can be written as
\begin{equation}\label{eq:DletaPhi}
\Delta\phi=-\int{\frac{ a^2}{a^2+b^2}d\theta}
\end{equation}
The independence of Eq.~(\ref{eq:DletaPhi}) with respect to the time
parametrization means that the rotation is $\emph{geometric}$
\cite{GeometricPhases}: the total rotation is independent of the
speed at which $a$ and $\theta$ change; the geometry of the change
-- the curve in $(a,\theta)$ space representing the change --
uniquely determines the total rotation.

Eq. (\ref{eq:DletaPhi}) can be written in the form
\begin{equation}\label{eq:LineIntFree}
\Delta\phi=-\oint\limits_\gamma{\frac{ a^2}{a^2+b^2}\,d\theta}+0\,da
\end{equation}
where $\gamma$ is the path in $(a,\theta)$ of the changes in the
shape of the cat. This has the form of a line integral
$\oint\limits_\gamma{\vec{A} \cdot d\vec{\ell}}$, where
$\vec{A}=\left(\frac{ a^2}{a^2+b^2},0\right)$ and
$d\vec{\ell}=\left(d\theta,da\right)$. By Stokes' theorem, we can
rewrite Eq.~(\ref{eq:LineIntFree}) as:
\begin{equation}\label{eq:SurfIntFree}
\Delta\phi=\int\limits_{S(\gamma)}{\mathcal{F}(a,\theta)\,da\,d\theta} =
\int\limits_{S(\gamma)}{\frac{2ab^2}{(a^2+b^2)^2}\,da\,d\theta}
\end{equation}
where $S(\gamma)$ is the oriented surface bounded by the path $\gamma$ and $\mathcal{F}=\frac{\partial A_\theta}{\partial a} - \frac{\partial A_a}{\partial\theta}=\frac{2ab^2}{(a^2+b^2)^2}$. Eq.~(\ref{eq:SurfIntFree}) tells us that the cat can indeed rotate itself: the change in $\phi$ for a closed path is the surface integral of $\mathcal{F}$ inside the path. Since $\mathcal{F}(\theta,a)\neq0$,  the rotation is not zero for a general path, and a cat changing its shape $(a,\theta)$ in the manner described by the path will rotate. We can also calculate the maximal angle this cat can rotate in one "stroke" (a simple closed path in $(a,\theta)$): since $0<\theta<\pi$ and $0<a<\infty$, the maximal rotation is given by $\int\limits_{0}^{\pi}{d\theta\int\limits_0^{\infty}{\frac{2ab^2}{(a^2+b^2)^2}\,da}}=\pi$. As we show in
section \ref{sec:TopoInv}, this is a consequence of the topology of the shape space.
%

\section{The Cat as a Gauge Problem}\label{section:gauge}

\begin{figure}
  \includegraphics[width=10 cm]{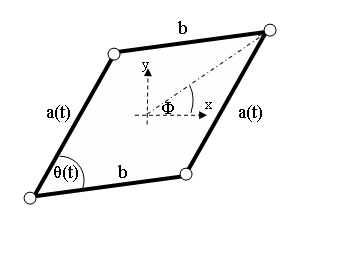}\\
  \caption{\textit{Example 2--} The cat model, $\phi$ chosen as the angle between $x$ and one of the masses.}\label{fig:Exmp2}
\end{figure}

Gauge theories are an important part of the formalism of fundamental physical laws. They are also known to play an important part in classical mechanics \cite{Lopez-gauge,GaugeMechanics}, swimming at low Reynolds number \cite{GeometricPhases}, and many other mathematical and physical problems. One such problem is the rotation of deformable bodies, as discussed in detail in \cite{GaugeMechanics,GeometricPhases,Montgomery}. As our model, being 2-dimensional with only two controls ($a$ and $\theta$) and one response($\phi$), is extremely simple, it is a suitable example for the construction of a gauge theory.

We will start by considering the problem we have already solved: the calculation of the total rotation of our system for a given sequence of changes in the "controls" $a$ and $\theta$. The total rotation
$\Delta\phi$ for a "stroke" is well-defined and easy to calculate using Eq.~(\ref{eq:SurfIntFree}).
The answer cannot be different for different ways of measuring $\phi$ because the system ends in the same shape in which it began; thus $\Delta\phi$ must be equal for all choices. But for a sequence of
changes that does not end with the same $a$ and $\theta$ as the initial configuration (i.e., not a "stroke"), the total rotation is not well-defined and depends on the way we have decided to characterize the angle $\phi$.

To demonstrate this, let us consider two cases (Figs.~\ref{fig:Exmp1}, \ref{fig:Exmp2}): in the first case, \textit{(Example 1)} $\phi$ is chosen as we chose it in section \ref{section:Model} - the angle between the $x$ axis of our reference frame and the $b$ rods. In the second case \textit{(Example 2)}, the angle $\phi$ is chosen as the angle between the $x$-axis and the line connecting two opposite masses. The two different choices for $\phi$ are legitimate: in both, $(a,\theta)$ and $\phi$ completely describe the system. By no \emph{a priori} consideration is one of these choices superior to the other (or any of the other ways to choose $\phi$). However, it is easy to see that for a path in $(a,\theta)$-space that is not closed,
$\Delta\phi$ is different in the two systems. For example, consider a change in $a$ alone while keeping
$\theta=\frac{\pi}{2}$: if $\phi$ is chosen as the angle of the $b$ rods with the $x$-axis, it will not change when $a$ changes at constant $\theta$ (and indeed - we have seen that $A_a$=0 in this system),
but if $\phi$ is defined as the angle of the line between the opposite masses, it will change even when $a$ alone changes.

We see that our system gives us the freedom to choose how to measure $\phi$. This freedom is called "gauge freedom." While some properties of the system are gauge-dependent ($\Delta\phi$ of non-closed paths; $\vec{A}$), others ($\Delta\phi$ of strokes; $\mathcal F$) are independent of gauge. This is analogous to the gauge freedom of the magnetic potential vector: the gauge-independent properties of magnetism are the magnetic field $\vec B$ (similar to $\mathcal F$ of the ''cat") and the total magnetic flux $\Phi$ through a closed loop $\gamma$ (corresponding to $\Delta\phi$). The gauge dependent properties are the potential vector $\vec A$ (an analog to $\vec A$ of the ''cat") and its integral over a non-closed path (analog to $\Delta\phi$). While in the case of magnetism it is difficult to understand what different gauges represent, as we cannot measure any gauge dependent quantity, in the case of the ''cat" different gauges are easily interpreted because gauge dependent quantities are readily calculable.
\section{Solving our Model in a Different Gauge}\label{sec:diffGauge}

The choice we made for $\phi$ in section \ref{section:Model} was not at all arbitrary; solving this problem directly in a different choice of $\phi$, as in \textit{Example 2}, can be very difficult - though it is clear that at least $\mathcal{F}$ must be the same in all cases. However, once we solve the system in one gauge, we can easily transfer our results to any new gauge without calculating $L$, which may be complicated. In order to do this, we can write $\phi$ in the new gauge (which we denote by $\tilde\phi$) as
\begin{equation}\label{eq:phiChange}
\tilde\phi=\phi + f(\theta,a)
\end{equation}
where $f(\theta,a)$ is the difference between the two measured angles (which is a function of $a$ and $\theta$, but not of $\phi$ itself). This implies that the infinitesimal change in $\tilde\phi$ can be expressed as
\begin{equation}\label{eq:infPhi}
d\tilde\phi=d\phi + \frac{\partial f}{\partial\theta}d\theta + \frac{\partial f}{\partial a}da = d\phi + \nabla f\cdot d\vec\ell
\end{equation}
where $\nabla=(\frac{\partial}{\partial\theta},\frac{\partial}{\partial a})$. Since $d\phi = \vec{A}\cdot d\vec\ell$, we can rewrite Eq.~(\ref{eq:infPhi}) as
\begin{equation}\label{eq:infTildePhi}
d\tilde\phi = (\vec{A} + \nabla f)\cdot d\vec\ell=\vec{\tilde A}\cdot d\vec\ell
\end{equation}
Thus, the new $A$ ($\vec{\tilde A}$) under the gauge transformation $\tilde\phi=\phi + f(\theta,a)$ can be expressed as
\begin{equation}\label{eq:GaugeTrans}
\vec{\tilde A} = \vec{A} + \vec \nabla f
\end{equation}
$\mathcal{F}$ will not change due the gauge transformation, since $\tilde{\mathcal{F}} = \frac{\partial\tilde A_\theta}{\partial a} - \frac{\partial\tilde A_a}{\partial\theta} = \mathcal{F} + \frac{\partial^2 f}{\partial\theta\partial a}-\frac{\partial^2 f}{\partial a \partial\theta}=\mathcal{F}$.
We can now easily calculate $\vec{A}$ for \textit{Example 2}: from a geometric calculation it is clear that
\begin{equation}\label{eq:ChangingGauge}
\tilde\phi=\phi + \arcsin\frac{a\sin\theta}{\sqrt{a^2+b^2+2ab\cos\theta}}
\end{equation}
thus
\begin{equation}\label{eq:AdiffGauge}
\vec{\tilde{A}}=(\frac{a^2}{a^2+b^2}+\frac{a\left(\cos\theta(a^2+b^2)+ab(1+\cos^2\theta)\right)}{(b+a\cos\theta)(a^2+b^2+2ab\cos\theta)},\frac{b\sin\theta}{\left(b^2+a^2+2ab\cos\theta\right)})
\end{equation}
In this gauge, as we have noted before, $A_a$ is nonzero.

To check this result, we can recalculate $\mathcal{F}=\frac{\partial A_\theta}{\partial a} - \frac{\partial A_a}{\partial\theta}=\frac{2ab^2}{(a^2+b^2)^2}$, which is identical to Eq.~(\ref{eq:SurfIntFree}), as it must be.

\section{The Importance of Non-Commutativity}\label{sec:non-commute}
\begin{figure}
  \includegraphics[width=10 cm]{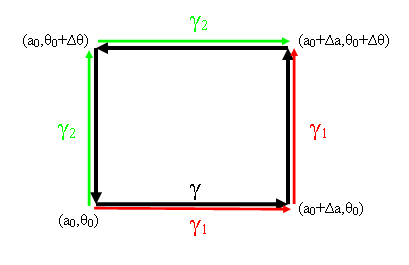}\\
  \caption{$\int\limits_\gamma{\vec{A}\cdot\vec{dl}}= \int\limits_{\gamma_1}{\vec{A}\cdot\vec{dl}}-\int\limits_{\gamma_2}{\vec{A}\cdot\vec{dl}}$}\label{fig:integral}
\end{figure}
Consider the same system, but now fix $\theta=\frac{\pi}{2}$ and let $b$ be a control instead. Clearly the system is now unable to rotate itself with zero angular momentum, though if we choose $\phi$ as in \textit{Example 2}, $\phi$ will change when changing only $a$. This implies that the ability to change orientation with zero angular momentum is measurable only on closed paths in shape-space -- which is equivalent to saying that a body can rotate itself if and only if $\mathcal{F}\neq0$.

\begin{figure}
  \includegraphics[width=10 cm]{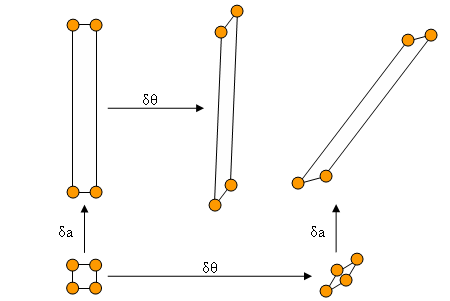}\\ \caption{Changing $a$ and $\theta$ in different orders leads to different rotations - this can easily be seen at large strokes.}\label{fig:LargStroke}
\end{figure}

Also, let us consider an infinitesimal rectangular path $\gamma=(a_0,\theta_0)\rightarrow(a_0,\theta_0+\Delta\theta)\rightarrow(a_0+\Delta a,\theta_0+\Delta\theta)\rightarrow(a_0+\Delta a,\theta_0)\rightarrow(a_0,\theta_0)$ (see Fig.~\ref{fig:integral}). If $\mathcal{F}\neq0$, the path integral $\oint\limits_\gamma{\vec{A}\cdot\vec{d\ell}}$  is nonzero. We can divide the path into two paths, $\gamma_1=(a_0,\theta_0)\rightarrow(a_0,\theta_0+\Delta\theta)\rightarrow(a_0+\Delta a,\theta_0+\Delta\theta)$ and $\gamma_2=(a_0,\theta_0)\rightarrow(a_0+\Delta a,\theta_0)\rightarrow(a_0+\Delta a,\theta_0+\Delta\theta)$, and write $\oint\limits_\gamma{\vec{A}\cdot\vec{d\ell}}=\int\limits_{\gamma_1}{\vec{A}\cdot\vec{d\ell}}-
\int\limits_{\gamma_2}{\vec{A}\cdot\vec{d\ell}}\neq0$, which implies that $\int\limits_{\gamma_1}{\vec{A}\cdot\vec{d\ell}}\neq\int\limits_{\gamma_2}{\vec{A}\cdot\vec{d\ell}}$ -- the two paths lead to different rotations! This means that the \emph{order} in which the deformations are made is important for bodies that can rotate with zero angular momentum, and vice-versa: in order for two deformations (followed by their inverses) to generate a net rotation, they must be non-commutative. In the case of the cat, changing $a$ and $b$ with fixed $\theta$ are two commutative operations -- the order in which they are made does not matter, and thus they do not generate a rotation. But changing $a$ and changing $\theta$ are non-commutative operations! This can be understood by comparing too large deformations: take the initial configuration to be a square ($a=b, \theta=\frac{\pi}{2}$) and change $a$ greatly. This will make our system very long and thin rectangle - a \emph{needle}. Then letting $\theta\rightarrow0$ will lead to only small rotation of the long axis of the needle. However, doing the changes in an opposite order will lead to a needle rotated by almost $\frac{\pi}{4}$ radians with respect to the first one (Fig.~\ref{fig:LargStroke}).

\section{A Cat in Honey - Aristotelian Mechanics}\label{sec:CatInHoney}
A "cat" is a deformable body which can change its orientation just by deforming itself. In a sense, this is an analog for a swimmer in a fluid, which is a deformable body that can change both its orientation and its position just by deforming itself. Thus, it is natural to ask how the "cat" will behave not just in empty space, but in some fluid. While in most cases this question is a very complicated one and shares very little with the empty space case, there is one regime in which the two questions share a lot - at low Reynolds number.

The dimensionless Reynolds number of a body in a fluid is defined the ratio of the inertial forces on the body to the viscous forces. At low Reynolds number the viscous forces dominate; the Navier-Stokes equations simplify to the time-invariant Stokes equations and inertia becomes irrelevant \cite{Purcell}.

Our "cat" is perfectly able to swim at low Reynolds number \cite{Stone}; just as in free space, the
"cat" is able to rotate but unable to translate itself. However, the calculation of its rotation in this case is significantly different from the previous calculation in free space. At $Re\ll1$, Newton's first law does not hold; force is instead proportional to velocity -- the system follows the physics of Aristotle, who believed that a body with no force acting upon it will not move. This regime also displays a characteristic of the "cat" in free space which was noted in section \ref{section:Model}: all motion is geometric, because the governing equations are independent of time.

Consider the case where the radii of the spheres are small relative to the distances between them, and
we can neglect the hydrodynamic interactions between the spheres (the interaction decays with $\frac{r}{x}$ where $r$ is the radii and $x$ is the distance \cite{Happel-Brenner}). In this case, the only force is the viscous force of the fluid on each sphere, which is of the form
\begin{equation}
\vec{F}_i = -6\pi\eta r_i\vec{v}_i
\end{equation}
where $\eta$ is the fluid viscosity and $r_i$ and $\vec{v}_i$ are the radius and velocity of the $i$'th sphere, respectively (we will assume that all spheres have the same radius $r_i=r$). There are no external forces or torques, so $\sum\vec{F}_i = \vec{0}$ and $\sum\tau_i = 0$.
From the symmetry of the "cat" it is clear that the former equation will be satisfied if and only if
the cat does not translate. We use $\sum\tau_i = 0$ as the equation of motion and calculate the rotation of the "cat" at low $Re$ for a given stroke as we did for the cat in free space.

The sum of the torques on each sphere is given by $6\pi\eta r \sum_i \vec{x}_i\times\vec{v}_i$, which
can be thought of as a scalar because the system is two-dimensional. But $\sum_i \vec{x}_i\times\vec{v}_i$ is just the angular momentum divided by the mass! Thus we can write $\sum\tau = 0 = L \left(\frac{6\pi\eta r }{m}\right)$. This is equivalent to the governing equation of the "cat" in free space, $L=0$, and all the calculations -- $A$, $\mathcal{F}$, etc. -- are identical! We have the surprising result that for the "cat", the motion at low Reynolds number without interactions is the same as the motion in free space. The reason for this can easily be seen from the equations: in free space, the equation of motion is $L=0$, and at low Reynolds number we have $\tau_{net} = 0$ -- because momentum in Newtonian mechanics maps to force in Aristotelian mechanics, these equations are the same. For us, this was a nice surprise, since the governing physics is completely different: for example, the "cat" does not dissipate energy in free space, but it does at low $Re$.

When the ratio of the distance between the spheres to their radii is not very large, we cannot neglect the hydrodynamic interactions between the spheres. In this case the expression for the force on each sphere is far more complicated, and is affected by the velocities of the other spheres as well. This means that there is no longer a simple map between the torque at low $Re$ and the angular momentum in empty space, and generally the two cases are different. However, the empty space and low $Re$ case still have much in common; for example, the time parametrization independency which enables us to calculate the total rotation through $\mathcal{F}(a,\theta)$. 
The ''cat" in free space and at low $Re$ number share several other properties as well -- for example, the integral of $\mathcal{F}$ over the entire shape space is $\pi$ in both cases -- the reason for which is discussed in the next section.

\section{Topological Invariants of the Cat}\label{sec:TopoInv}
We showed earlier that the total rotation of the maximal stroke is $\pi$. Interestingly, this is the case for our "cat" in both free space and at low Reynolds number -- even when one takes into account the interactions between the four spheres. The interactions in general change $\mathcal{F}$, but not the integral of $\mathcal{F}$ over the whole shape space; this is a nice example of a topological invariant which characterizes our system \cite{TopoInvar,Nakahara}. To demonstrate this, we will consider the case in which $\theta$ can take values up to $2\pi$ and not just $\pi$. If we show that in this case the rotation is $2\pi$, it is clear from symmetry that in the case $\theta\leq\pi$ the total rotation must be $\pi$.

Let us look at the case $a=b=1$ for simplicity, and change $\theta$ from 0 to $2\pi$, keeping $a=1$. Since in this case $\vec A=\left(\frac{1}{2},0\right)$, we get $\Delta\phi=\pi$. However, trying to do the same calculation using $F$ instead of $\vec A$ we face a problem: what area does this curve bound? We can look at it in two different ways:         
\begin{itemize}
\item We can complete our stroke by changing $a$, at $\theta=2\pi$, from 1 to 0, then changing $\theta$ to 0 (keeping $a=0$), and finally changing $a$ back to $1$. The changes in $a$ will not change $\Delta\phi$ because $A_{a}=0$; even if this is not the case (if we use a different gauge), the two paths (increasing and decreasing $a$) must cancel each other because $\theta$ is the same on both. The changes in $\theta$ do not change $\Delta\phi$ either: at $a=0$ we have $\vec A_\theta =0$. Thus $\Delta\phi$ must be the same in the complete and incomplete strokes. The integral of $\mathcal{F}$ over this area (which we denote as $S_1$) gives $\pm\pi$ (the sign depends on choice of the surface orientation).
\item Alternatively, we can change $a$ (at $\theta=2\pi$) to infinity (instead of 0), then change $\theta$ to zero, and then change $a$ back to 1. At $a=\infty$, $A_\theta=0$ just as at $a=0$, so again all the added parts of the stroke will not change $\Delta\phi$. The integral of $\mathcal{F}$ over this area (which we denote as $S_2$), is $\mp\pi$. The sign is opposite,
    since the two strokes bound areas with opposite orientations!
\end{itemize}
We conclude that the two ways of calculating $\Delta\phi$ are not necessarily the same, but they must only differ by $2\pi n$ for some integer $n$ (that is $\int\limits_{S_1}{\mathcal{F}da}-\int\limits_{S_2}{\mathcal{F}da}=2\pi n$), and in our case $n=1$. But since the orientations of $S_1$ and $S_2$ are opposite, the difference between the integrals is just the integral of $\mathcal{F}$ over the two areas with the same orientation.  That is to say $\int{\mathcal{F}da}=2\pi n$ (the integral is taken over all shape space), for some integer $n$, which in our case turns out to be 1. Thus the integral of $\mathcal{F}$ over all shape space can only be changed in $2\pi$ ''jumps", and cannot take any value.

\begin{figure}
  \includegraphics[width=10 cm]{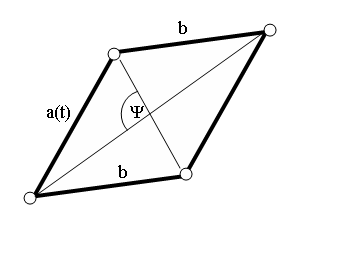}\\
  \caption{Instead of $a(t)$, one can use $\psi$ to define the shape}\label{fig:Cat_psi}
\end{figure}

The  "trick" of going through $a=0$ or $a=\infty$ is not really necessary. If we use a
different coordinate system: suppose we choose $\psi$ -- the angle between the center of mass and the two masses along the same $a$ rod -- as a control instead of $a$ (see Fig \ref{fig:Cat_psi}). In this case, $a=0 \Leftrightarrow \psi=0$ and $a=\infty \Leftrightarrow \psi=\pi$. In this parametrization our shape space looks like the surface of a sphere: it has two coordinates $(\theta,\psi)\in[0,2\pi)\times[0,\pi]$. At both $\psi=0$ and $\psi=\pi$, $\theta$ loses its meaning, just like the longitude coordinate on a sphere.
Now instead of adding paths, we can just integrate over the surface that includes the poles. Since each closed loop on the sphere bounds two areas with opposite signs, the rest of the argument is the same.

The fact that the total curvature of our system can only "jump" in $2\pi$ steps teaches us something about the system: for example, we see immediately that the total curvature will still be $2\pi$ when there are different masses on the two diagonals: starting with equal masses and changing them slowly, $\mathcal{F}$ must change continuously at each point, so the total integral over $\mathcal{F}$ must also change continuously. However, we know it can only change in $2\pi$ jumps, so it does not change at all! This also explains why the interactions between the spheres at low Reynolds number, as we clammed, do not change the total curvature either: for infinitesimal spheres, where the interactions are negligible, the system acts just like the free space cat, and the total curvature is also $2\pi$. Inflating the spheres continuously cannot change the total integral over $\mathcal{F}$, although the interactions are not negligible anymore!

\paragraph{ Acknowledgments} We thank SciTech 2007 for giving us the opportunity to do this research and  J.E Avron for discussions. O.R is supported by the ISF and the fund for promotion of research at the Technion.


\end{document}